\documentclass[12pt]{article}

\usepackage[letterpaper,hmargin=1in,vmargin=1in]{geometry}

\usepackage{graphicx,epstopdf,amsmath,amsfonts,amssymb,sectsty,cite}

\def\be{\begin{equation}}
\def\ee{\end{equation}}
\def\ba{\begin{eqnarray}}
\def\ea{\end{eqnarray}}
\def\ge{\mathrel{\raise.3ex\hbox{$>$\kern-.75em\lower1ex\hbox{$\sim$}}}}
\def\la{\mathrel{\raise.3ex\hbox{$<$\kern-.75em\lower1ex\hbox{$\sim$}}}}

\def\simgt{\mathrel{\raise.3ex\hbox{$>$\kern-.75em\lower1ex\hbox{$\sim$}}}}
\def\simlt{\mathrel{\raise.3ex\hbox{$<$\kern-.75em\lower1ex\hbox{$\sim$}}}}

\newcommand{\bi}[1]{\bibitem{#1}}
\newcommand{\fr}[2]{\frac{#1}{#2}}

\newcommand{\nc}{\newcommand}

\nc{\gone}{\bar g_{\pi NN}^{(1)}}
\nc{\gzero}{\bar g_{\pi NN}^{(0)}}
\nc{\al}{\alpha}
\nc{\ga}{\gamma}
\nc{\de}{\delta}
\nc{\ep}{\epsilon}
\nc{\ze}{\zeta}
\nc{\et}{\eta}
\nc{\ka}{\kappa}
\nc{\rh}{\rho}
\nc{\si}{\sigma}
\nc{\ta}{\tau}
\nc{\up}{\upsilon}
\nc{\ph}{\phi}
\nc{\ch}{\chi}
\nc{\ps}{\psi}
\nc{\om}{\omega}
\nc{\Ga}{\Gamma}
\nc{\De}{\Delta}
\nc{\La}{\Lambda}
\nc{\Si}{\Sigma}
\nc{\Up}{\Upsilon}
\nc{\Ph}{\Phi}
\nc{\Ps}{\Psi}
\nc{\Om}{\Omega}
\nc{\ptl}{\partial}
\nc{\del}{\nabla}
\nc{\ov}{\overline}
\nc{\newcaption}[1]{\centerline{\parbox{15cm}{\caption{#1}}}}
\nc{\us}{U(1)$_S$}

\def\beq{\begin{equation}}
\def\eeq{\end{equation}}
\def\bmat{\begin{displaymath}}
\def\emat{\end{displaymath}}
\def\bear{\begin{eqnarray}}
\def\eear{\end{eqnarray}}
\def\ba{\begin{eqnarray}}
\def\ea{\end{eqnarray}}
\def\bery{\begin{array}}
\def\ery{\end{array}}
\def\bit{\begin{itemize}}
\def\eit{\end{itemize}}
\def\ben{\begin{enumerate}}
\def\een{\end{enumerate}}
\def\btab{\begin{tabular}}
\def\etab{\end{tabular}}
\def\btbl{\begin{table}}
\def\etbl{\end{table}}
\def\bfig{\begin{figure}[htb]}
\def\efig{\end{figure}}
\def\bpic{\begin{picture}}
\def\epic{\end{picture}}

\def\ga{\mathrel{\raise.3ex\hbox{$>$\kern-.75em\lower1ex\hbox{$\sim$}}}}
\def\la{\mathrel{\raise.3ex\hbox{$<$\kern-.75em\lower1ex\hbox{$\sim$}}}}
\def\gappeq{\mathrel{\rlap {\raise.5ex\hbox{$>$}}
{\lower.5ex\hbox{$\sim$}}}}
\def\lappeq{\mathrel{\rlap{\raise.5ex\hbox{$<$}}
{\lower.5ex\hbox{$\sim$}}}}

\def\gyr{{\rm \, G\kern-0.125em yr}}
\def\mev{{\rm \, Me\kern-0.125em V}}
\def\gev{{\rm \, Ge\kern-0.125em V}}
\def\tev{{\rm \, Te\kern-0.125em V}}

\sectionfont{\large}
\subsectionfont{\normalsize}

\begin{document}

\begin{titlepage}

\setcounter{page}{1}

\vspace*{0.2in}

\begin{center}

{\LARGE \bf  Observing a light dark matter beam  with neutrino experiments}

\vspace*{1cm}
\normalsize

\vspace*{0.5cm}
\normalsize

{\bf Patrick deNiverville$^{\,(a)}$, Maxim Pospelov$^{\,(a,b)}$, and Adam Ritz$^{\,(a)}$}

\smallskip
\medskip

$^{\,(a)}${\it Department of Physics and Astronomy, University of Victoria, \\
     Victoria, BC, V8P 5C2 Canada}
     
     $^{\,(b)}${\it Perimeter Institute for Theoretical Physics, Waterloo,
ON, N2J 2W9, Canada}

\smallskip
\end{center}
\vskip0.2in

\centerline{\large\bf Abstract}

We consider the sensitivity of fixed-target neutrino experiments at the luminosity frontier to light stable states, such as those present in models of MeV-scale dark matter.
To ensure the correct thermal relic abundance, such states must annihilate via light mediators, which in turn provide an access portal for direct production in colliders or fixed targets. Indeed, this framework endows the neutrino beams produced at fixed-target facilities with a companion  `dark matter beam', which may be detected via an excess of elastic scattering events off electrons or nuclei in the (near-)detector. We study the high luminosity proton fixed-target experiments at LSND and MiniBooNE, and determine that the ensuing sensitivity to light dark matter generally surpasses that of other direct probes. For scenarios with a kinetically-mixed U(1)$'$ vector mediator of mass $m_V$, we find that a large volume of parameter space is excluded for $m_{\rm DM}\sim 1-5$~MeV, covering vector masses  $2 m_{\rm DM} \la m_V \la m_\eta$ and a range of kinetic mixing parameters reaching as low as $\ka \sim 10^{-5}$. The corresponding MeV-scale dark matter scenarios motivated by an explanation of the galactic 511 keV line are thus strongly constrained.

\vfil
\leftline{July 2011}
    
\end{titlepage}

\section{Introduction}

While the empirical evidence for dark matter (DM), through its gravitational effects in astrophysics and cosmology, derives from many sources and ranges over many distance scales, the search for any signature of its non-gravitational interactions remains one of the focal points of research in particle physics. 
Thermal relic weakly-interacting massive particles (WIMPs), predicted or otherwise introduced into many extensions of the Standard 
Model (SM), represent an appealing dark matter candidate. In particular, though the WIMP mass scale and couplings are only weakly constrained 
by the requirement of the correct relic abundance, the characteristic weak-scale parameters of the paradigmatic WIMP 
fall into a range that offers hope for the direct discovery of non-gravitational DM interactions in the laboratory.
However, in recent years, considerable attention has been paid to particle physics models that deviate from the minimal idea of a single WIMP with 
weak-scale interactions with the SM. Possibilities of both light dark matter candidates 
and/or light mediator particles have been explored  \cite{Boehm,Fayet,BF,light-chi,Neil,PRV,HZ,AFSW,PR,Drees,tests,BPR,slac,Reece,best,bpr99c,others} 
with the motivation of tying various anomalous experimental signatures to the annihilation, scattering or decay of dark matter. 
While most anomalies will likely find other explanations, the expanded mass range for  WIMP candidates and mediators opens 
a number of new experimental avenues, which go beyond the characteristic direct detection strategies for a 
minimal (weak-scale) WIMP.

In this paper, we revisit a class of MeV-scale dark matter models, 
originally designed to explain the unusual strength and morphology 
of the 511 keV emission observed from the galactic center with annihilating dark matter \cite{Boehm}.
MeV-scale models of thermal relic dark matter require the existence of a light mediator \cite{Fayet,BF,light-chi,Neil,PRV,HZ,AFSW,PR}, so that cosmological freeze-out occurs as a rescaled version of conventional WIMP freeze-out with reduced mass and temperature scales.  The existence of light mediators, and thus a more complex light hidden (or dark) sector, naturally stimulates interest in the low and intermediate energy particle physics manifestations \cite{Fayet,Drees,tests,BPR,slac,Reece}. Certain classes of flavor-conserving light states with  lifetimes below 1 second are often immune to a variety of astrophysical, cosmological, and collider tests, even if their interactions are larger than the characteristic weak rate. Specifically, in scenarios where states in the hidden sector are in the hadronic mass range, and have a lifetime longer than other hadronic states which undergo weak decays, fixed target experiments with detectors 10--1000m from the target, as in modern long-baseline neutrino experiments, can provide complementary sensitivity to colliders \cite{best,bpr99c,others}. 
A rather striking consequence of models with light (sub-GeV) dark matter is the production of a high intensity `dark matter beam', generated as dark matter particles are pair-produced as a result of the proton-target interactions and boosted along the proton beam direction \cite{bpr99c}. The scattering of light dark matter in the (near-)detector would then generate an additional source of neutral-current-type scattering events (see, {\em e.g.} \cite{losecco}). This prediction implies that a direct search for MeV-scale stable dark matter is possible at experiments at the luminosity frontier, which is the focus of the present paper.

To motivate the importance of fixed target facilities in this low mass regime, we recall \cite{bpr99c} that given such a hidden sector, assumed neutral with respect to the SM gauge group, we can parametrize the interactions as follows,
\be
\label{int}
{\cal L}_{\rm mediation} = \sum_{d_1,d_2} \fr{{\cal O}^{(d_1)}_{\rm  NP} {\cal O}^{(d_2)}_{\rm SM} }{\Lambda^{d_1+d_2-4}},
\ee
where ${\cal O}$ denotes SM and new physics (NP) operators of canonical dimensions $d_1$ and $d_2$, and $\Lambda$ is a cutoff scale presumably at a TeV or above. Light, long-lived, hidden sector states can be studied at high-luminosity fixed target experiments, where the production cross-section mediated by an interaction (\ref{int}) of 
dimension $d_1+d_2=4+n$ (with $n\geq 0$) typically scales as $\si \sim E^{2n-2}/\La^{2n}$. 
Inserting typical numbers for the attainable luminosities and typical energies at high-energy colliders and proton fixed-target machines respectively, leads to an interesting comparison in the total production count (denoted $N$) of neutral GeV-scale states  \cite{bpr99c}:
\be
\fr{N_{\rm collider}}{N_{\rm target}} 
\sim 10^{-12 + 6n}.
\label{counting}
\ee
 It is apparent that for a marginal interaction, $n=0$, the production rates at fixed targets may be sufficiently advantageous to easily counteract the low geometric acceptance of a detector placed some distance from the target, and such facilities can provide the dominant level of experimental sensitivity.  The set of relevant or marginal interactions forms a small, but generically the most important, subset of interactions in (\ref{int}) known as SM portals \cite{pw,portal1,higgsportal,DMportal}:
\be
{\cal O}^{n\leq 0}_{\rm SM}=F^Y_{\mu\nu}, H^\dagger H, LH,  \label{portals} 
\ee
where $F^Y_{\mu\nu}$, $ H$ and $L$ are the hypercharge field strength, and the Higgs and lepton doublets. The operators (\ref{portals}), denoting respectively the {\it vector}, {\it Higgs}, and {\it neutrino} portals, allow a coupling of the SM to (SM neutral) new physics at the renormalizable level.

This paper aims to explore the sensitivity of existing (and future) experimental infrastructure for long-baseline neutrino experiments to light dark matter which forms part of a hidden sector interacting with the SM through the portals (\ref{portals}). At fixed targets, these interactions entail  the production (along with the neutrinos) of a boosted dark matter beam through the generation and subsequent decay of GeV-scale mediators. As we will show below, existing data from high-luminosity experiments such as LSND and MiniBooNE already imposes stringent constraints on viable scenarios of MeV-scale dark matter, due to the limits on neutral-current-like scattering events off electrons and nuclei in the detector. While the idea of searching for exotics using fixed target facilities is certainly not new (see e.g. \cite{axions_exp,neutrino_exp,gluino_exp,unstable, losecco}), long-baseline neutrino facilities introduce a particular advantage for probing stable states in that the large mass of the (near-)detector can be utilized to observe scattering rather than just the results of a decay. We will argue that these facilities already provide the dominant constraints on many models of this type, and specifically those utilizing the vector portal ${\cal O}_{\rm SM} = F^Y_{\mu\nu}$ which are generally less constrained in other ways. In particular, for scenarios of MeV-scale dark matter which aim to explain the galactic 511~keV line, these constraints 
are generically more stringent than those derived from rare meson decays \cite{Fayet}. 
 
The rest of this paper is organized as follows. In section \ref{sec:Model}, we describe and motivate one of the most viable classes of MeV-scale hidden-sector dark matter, which   interacts with the SM via the vector portal, and the parameter constraints from astrophysical and cosmological data. In section \ref{sec:DMProd}, we explain how the parameter space of this model may be probed at fixed target neutrino oscillation experiments and follow this with an analysis of the sensitivity of the LSND and MiniBooNE experiments. We conclude by contrasting this sensitivity  with other limits that can be placed on MeV scale states, and explore possibilities for future progress, in section \ref{sec:Conc}.\footnote{[Note Added - August 12, 2013] The update includes a short section at the end, showing additional sensitivity plots, which aims to clarify the 90\% exclusion contours on the parameter space of the vector portal model from LSND's elastic scattering analysis \cite{lsnd_elastic}.}

\section{Light MeV-scale thermal relic dark matter}
\label{sec:Model}

The viability of thermal relic dark matter with a mass in the MeV--GeV range, well below the Lee-Weinberg bound,  seemingly rests on the presence of a light hidden sector with states which can mediate annihilation \cite{BF,Fayet,PRV}. Moreover, various phenomenological constraints \cite{PRV} suggest that the most viable scenarios are those in which the hidden sector is uncharged under Standard Model symmetries. This naturally leads us to the portal interactions (\ref{portals}), as the primary means of probing these sectors at low energies.

For a thermal relic dark matter (TRDM) candidate in the MeV mass range, the dominant decay channels will lead to $e^+e^-$, with direct annihilation to photons and neutrinos often suppressed (a reduced coupling to neutrinos being a phenomenological  constraint to eliminate drastic softening of the supernova neutrino spectra).  The fact that annihilation of light TRDM generically produces positrons naturally implies that galactic observations are a significant source of constraints. Indeed, MeV-scale models were initially motivated by the 511~keV line observed from the galactic centre 
\cite{Boehm}, and recently mapped out in considerable detail by INTEGRAL/SPI \cite{integral,bouchet}. However, it is important to emphasize that, independent of any attempt to explain its source, the magnitude of this flux provides quite a significant constraint on light dark matter models in this class. To see this, we parametrize the flux $\Ph$ observed by INTEGRAL/SPI in the form,
\be
\label{eq:511}
 \frac{\Ph_{511,{\rm DM}}}{\Ph_{511,{\rm tot}}} \sim 10^4 N_{e^+} \times \frac{\langle \si v \rangle_{\rm gal}}{{\rm pbn}} \times \left(\frac{1\,{\rm MeV}}{m_{\rm DM}}\right)^2\times\left(\frac{\Om_{\rm DM}}{\Om_m} \right)^2,
\ee
where $N_{e^+}$ is the number of positrons per decay and $\Om_{\rm DM}/\Om_m \sim (1\,{\rm pbn})/\langle \si v\rangle_{\rm fo}$. The result depends crucially on the annihilation rates at freeze-out $\langle \si v\rangle_{\rm fo}$ and in the galactic centre $\langle \si v\rangle_{\rm gal}$. We see that if annihilation is dominantly to $e^+e^-$, any MeV-scale dark matter candidate with relic abundance close to $\Om_m$ should have an annihilation rate in the galaxy suppressed by several orders of magnitude relative to the rate at freeze-out. This is quite a strong constraint, and as emphasized in \cite{BF,PRV} tends to single out one class of models as the most viable. We summarize the issues below for the case of interactions mediated via the vector and Higgs portals,  with a light hidden sector containing a dark matter state and a U(1) mediator $V$ or singlet scalar mediator $S$ respectively \cite{PRV}:

\begin{itemize}
\item[(\checkmark)] {\it Vector portal, $m_{\rm DM} < m_V$}: A scalar DM candidate has $p$-wave annihilation which satisfies (\ref{eq:511}) since $v\sim 10^{-3}$ in the galaxy, and thus is viable for sub-percent mixing via the portal coupling.
\item[(X)] {\it Vector portal, $m_{\rm DM} > m_V$}: This implies $s$-wave annihilation, and thus the 511 keV flux limit (\ref{eq:511}) can only be satisfied with a highly subdominant component of dark matter.
\item[(X)] {\it Higgs portal, $m_{\rm DM} < m_S$}: Annihilation is suppressed, and would require {\cal O}(1) mixing via the Higgs portal which is ruled out for example by $K$ and $B$ decays.
\item[(X?)] {\it Higgs portal, $m_{\rm DM} > m_S$}: A fermionic DM candidate has $p$-wave annihilation which can satisfy (\ref{eq:511}), but needs a high degree of tuning to avoid limits on $K$ decays with missing energy.
\end{itemize}

This phenomenological analysis, described in more detail in \cite{PRV}, motivates models interacting via the vector portal as the most natural setting for light MeV-scale TRDM, with
the dark matter state being the lightest in the hidden sector. In the next subsection, we will outline this model in more detail and then move on to explore how it may be probed in fixed target experiments.

\subsection{Light dark matter with a U(1) mediator}
\label{ssec:Model}

We consider a hidden sector, charged under a U(1)$'$ gauge group, with a vector portal coupling to the SM via kinetic mixing (see {\it e.g.} \cite{PRV}). We also assume that the U(1)$'$ is spontaneously broken at a low scale by a Higgs$'$ sector, leading to a mass for the vector mediator $V_\mu$. The relevant low energy Lagrangian takes the form,
\begin{equation}
\label{eq:bmodlag}
 {\cal L}_{V,\chi}=-\frac{1}{4}V_{\mu \nu}^2+\frac{1}{2} m_{V}^2 V_\mu^2 +\kappa V_\nu \partial_\mu F^{\mu \nu} +|D_\mu \chi |^2 - m_\chi^2 |\chi|^2+{\cal L}_{h'},
\end{equation}
where $\chi$ is the complex scalar dark matter candidate, taken to be stable due to a suitable $\mathbb{Z}_2$ parity, and the U(1)$'$ covariant derivative is $D_\mu = \ptl_\mu + i e' V_\mu$.  All of the kinetic terms and interactions involving the Higgs$'$ are included in the ${\cal L}_{h'}$ term and will not play a role here, although we require that the full scalar potential is such that the physical Higgs$'$ is more massive than $\chi$. In general, $V$ mixes kinetically with the hypercharge gauge boson, but at low energies we can ignore the induced coupling to the $Z$, and in (\ref{eq:bmodlag}) we have rescaled the kinetic mixing parameter $\ka$ so that it just reflects mixing with the photon.  

\begin{figure}
\centerline{\includegraphics[viewport=60 630 550 720, clip=true, width=16cm,angle=0]{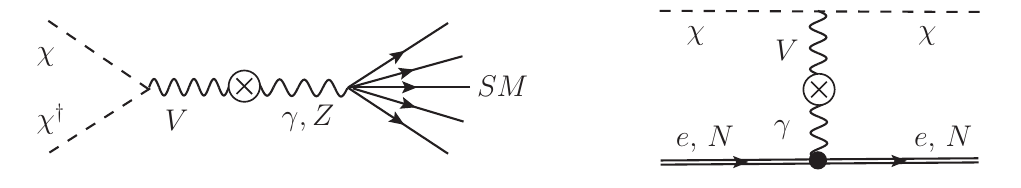}}
 \caption{\footnotesize  Tree-level annihilation (left) and scattering (right) of  scalar dark matter in the U(1)$'$ hidden sector.}
\label{f1} 
\end{figure}

The model contains four parameters; the masses $m_\ch$ and $m_V$ of the dark matter candidate and the vector mediator, the U(1)$'$ gauge coupling $e'$, and the kinetic
mixing coefficient $\ka$. On requiring that $\ch$ comprises the majority of dark matter, the constraint on its relic abundance allows us to fix one relation between these four parameters. The primary quantity here is the annihilation rate, which is given in general by the diagram on the left of Fig.~\ref{f1}. In practice, we will be in a regime here where the branching is predominantly to an $e^+e^-$ final state. In the limit of small mixing, and dropping a small correction proportional to $m_e^2$, the rate for annihilation is given by \cite{BF},
\be
\label{eq:annihln}
 \langle \si v\rangle_{\rm ann} \simeq 3 \times 10^{-27}\,{\rm cm}^2 \times \left(\frac{\ka^2\al'}{\al} \langle v^2\rangle \right) \times \left(\frac{\rm MeV}{m_\ch}\right)^2 \times \sqrt{1-\frac{m_e^2}{m_\ch^2}} 
 \left(\frac{4m_\ch^2}{4m_\ch^2-m_V^2}\right)^2.
\ee
Noting the $p$-wave suppression, with $v\sim 0.3$ at freeze-out,  we observe that the WMAP constraint on the relic density 
$\Om_{\rm DM}h^2 \sim 0.1 \sim (0.1\,{\rm pbn})/\langle \si v\rangle_{\rm fo}$ imposes the following restriction on the model parameters, 
\begin{equation}
 \label{eq:freeze}
 \frac{\alpha' \kappa^2}{\alpha} \times \left(\frac{10\, {\rm MeV}}{m_V} \right)^4 \times \left(\frac{m_\chi}{1\, {\rm MeV}} \right)^2 \sim 3 \times 10^{-6},
\end{equation}
where to simplify the presentation we have taken $m_e^2 \ll m_\ch^2 \ll m_V^2$, relations which are satisfied up to ${\cal O}(25\%)$ in the parameter regimes studied here. However, we use the more  precise constraints from (\ref{eq:annihln}) in the subsequent numerics, reducing the number of free parameters to three, which we will take to
be $\{m_\ch,m_V,\ka\}$. Note also that the $p$-wave suppression of annihilation for low velocities allows this process to satisfy the 511~keV flux constraint (\ref{eq:511}), as alluded to above, as well as the CMB constraints on dark matter 
annihilation \cite{CMBa}.

A rotation of the annihilation diagram describes scattering off electrons and nucleons as shown on the right of Fig.~\ref{f1}, and provides a means for detecting the presence of light DM in the galactic halo, at least in principle. However, in practice MeV-scale dark matter only has a characteristic  kinetic energy of ${\cal O}$(eV) 
in the Earth's rest frame, leading to a recoil in nuclear scattering which is well below the detection threshold for the current generation of underground direct detection experiments.
However, this problem could be circumvented if dark matter were first boosted to $v \sim 1$ before it undergoes scattering. Indeed, for sufficiently low masses, it is feasible to produce a {\it dark matter beam} in collider or fixed target experiments that may see MeV dark matter in the ultrarelativstic regime, $E_\chi \gg m_\chi$. In particular, modern long-baseline neutrino facilities are ideal for this purpose, having a high luminosity, and also large (near-)detectors, which can be used to search for dark matter elastic scattering.
 We will turn to this possibility, and the sensitivity that can be attained, in the next section.

\section{Production and scattering of the dark matter beam}
\label{sec:DMProd}

Within the hidden sector scenario outlined in the previous section, and for sufficiently small $m_V$, the following chain of processes can produce a dark matter beam at a fixed target experiment:

\begin{enumerate}
  \item $p+p \to X + \pi^0,\eta$
  \item $\pi^0,\eta \to \gamma + V$
  \item $V \to 2\chi$
\end{enumerate}

Depending on the beam energy and form of the target, the relevant decay lengths ensure that this entire sequence of events will occur either inside the target itself or in the subsequent decay volume. Depending on $m_V$, the dominant production mode will be $\pi^0$ or $\eta$ decays, and we have focused on this subset of hadronic states due to their large branching fraction to photons. While $\pi^0$'s dominate production for $m_V < m_\pi$, the addition of the $\eta$ mode allows access to a larger range in $m_V$, and consequently $m_\ch$.
 In both cases, the branching ratio to $V$ is proportional to that of the radiative decays of the mesons to two photons, though suppressed by $\kappa^2$ and phase space factors related to the ratio of $m_V$ to $m_\phi$ where $\phi=\pi,\eta$,
\begin{equation}
 \label{eq:brV}
 \textnormal{Br}_{\phi \to \gamma V} \simeq 2 \kappa^2 \left (1 - \frac{m_V^2}{m_\phi^2} \right)^3 \textnormal{Br}_{\phi \to \gamma\gamma}.
\end{equation}
For the case of $\pi^0$ decays $\textnormal{Br}_{\pi^0 \to \gamma\gamma}\simeq1$, while for $\eta$ decays $\textnormal{Br}_{\eta \to \gamma\gamma}\simeq0.39$.

Given that  we require $\ka \ll 1$, it follows that $V$ generically decays within the hidden sector, Br$_{V\rightarrow 2\ch} \simeq 1$, and the ensuing dark matter beam then propagates along with the neutrino beam. For the range of $\ka$ values considered here, it has a weak-scale scattering cross-section with normal matter  and may be detected  through neutral current-like processes, either with electrons  $e+\chi \to e + \chi$, or nucleons, $N+\chi \to N+\chi$. In order to probe this scenario, we will utilize the results of LSND and MiniBooNE, which have two of the largest datasets and importantly have published analyses on neutrino elastic scattering, which DM scattering will closely mimic. Note that due to their respective beam energies, both LSND and MiniBooNE are  sensitive to electron scattering, while only MiniBooNE is sensitive to elastic scattering off nucleons.

\subsection{Dark matter beams at LSND}
\label{ssec:DMLSND}

We now probe the parameter space of the model by calculating the number of dark matter neutral current-like elastic scattering events that would be expected at the LSND experiment, $N_{\textnormal{events}}$, and compare it to the total number of elastic (neutral and charged current) scattering events off electrons actually observed \cite{lsnd_elastic}. At LSND, pions were produced by impacting an 800 MeV proton beam onto either a water or high-$Z$ metal target \cite{lsndresult}. The LSND experiment provides the largest fixed-target sample of pions currently available, and has the potential to provide the most stringent limits on the model parameter space for the range of $m_V$'s to which it is sensitive.

The overall normalization of the event rate at LSND  is dictated by $N_{\pi^0}$, the total number of neutral pions produced over the lifetime of the experiment. 
In practice, we can approximate $N_{\pi^0}$ by equating the $\pi^0$ production rate with that of $\pi^+$, on the grounds that the measured $\pi^0$ and $\pi^+$ production rates in proton-nucleon collisions differ by ${\cal O}(1)$ factors (see eg. \cite{etacross}).  We estimate $N_{\pi^+}$ by working backwards from the neutrino flux reported by the collaboration, and as the majority of the neutrinos were products of $\pi^+$ decays at rest,
\begin{equation}
 \label{eq:Npi}
 N_{\pi^+} = \frac{\Phi_\nu \times A_{\textnormal{det}}}{(d\Omega_{\textnormal{lab}}/4\pi)_\nu} \approx 10^{22}.
\end{equation}
Here $\Phi_\nu=1.3\times 10^{14}$ $\nu$ cm$^{-2}$ is the neutrino flux over the lifetime of the experiment, $A_{\textnormal{det}}\simeq2.5\times 10^5$ cm$^2$ is the area of the detector facing the target, and $(d\Omega_{\textnormal{lab}}/4\pi)_\nu \approx 3\times 10^{-3}$ is the fraction of the solid angle subtended by the detector relative to the target.

We employed a Monte Carlo simulation to determine the dark matter flux incident on the LSND detector. Pions were generated in the momentum ranges expected by LSND over an array of possible angles. According to the appropriate branching fractions, the subsequent decays to $\pi^0\rightarrow V\gamma$ and  $V\rightarrow\chi\chi^\dagger$'s, were simulated and the trajectories of the $\chi$'s were then checked to determine if they intersected with the detector. A re-weighting technique was then used to weight each trajectory according to the momentum and angular distribution of the initial $\pi^0$. We assumed that this distribution was similar to the production distribution of $\pi^+$'s and used the parameterization of the production cross-section by Burman and Smith \cite{lsndpionpara}. It was also necessary to account for the fact that the pion production distribution was not constant throughout the lifetime of the experiment, as LSND made use of two different targets \cite{lsndresult}, and the 
resulting normalized (and azimuthally symmetric) distribution, which we denote $f_{\pi}^{\rm BS}(\theta,p)$, is a weighted average.

With the pion distribution in hand, the simulation determined the number of $\chi$'s which reach the detector, along with their energies and the distance travelled through the detector. To determine the expected number of elastic scattering events, we modelled the detector as a cylindrical tank filled with mineral oil CH$_2$ (see \cite{lsnddetector} for further details). The scattering cross section for $e \chi \to e \chi$, assuming $E \gg m_e$,
takes the form
\begin{equation}
 \label{eq:escatter}
 \frac{d\sigma_{e\chi\to e\chi}}{dE_f} = \frac{\alpha' \kappa^2}{\alpha} \times \frac{4 \pi \alpha^2 \left(2 m_e (E^2- EE_f)- m_\chi^2 E_f \right)}{E^2(m_V^2+2 m_e E_f)^2},
\end{equation}
where $E$ is the energy of the incoming dark matter particle and $E_f$ is the energy of the scattered electron. We use (\ref{eq:freeze}) to replace the ratio $\frac{\alpha' \kappa^2}{\alpha}$ with a function of $m_V$ and $m_\chi$. The number of elastic scattering events of dark matter off electrons can now be schematically represented as follows,
\begin{equation}
 \label{eq:Nevents}
 N^{\rm LSND}_{e\chi \to e\chi} = n_e \times N_{\pi^0} \times \textnormal{Br}_{\pi^0 \to \gamma V} \times \ep_{\rm eff} \times \sum_i L_i \sigma_{e\chi\to e\chi}(E_i) f_{\pi}^{\rm BS}(\theta_i,p_i) \Delta_i,
\end{equation}
where $n_e \simeq 5.1 \times 10^{23}\, \textnormal{electrons/cm}^{-3}$ is the number density of electrons in mineral oil, $\textnormal{Br}_{\pi^0 \to \gamma V}$ is calculated in (\ref{eq:brV}), and $\ep_{\rm eff}\simeq 0.19$ is the electron detection efficiency within LSND's elastic scattering analysis \cite{lsnd_elastic}, that we will use for
comparison. In order to map out the full production distribution, the remaining sum is over all simulated trajectories, where $L$ is the distance travelled through the detector and $\sigma_{e\chi\to e\chi}$ is the integrated cross section from (\ref{eq:escatter}) for outgoing electrons with recoil energies between $18$ MeV and $50$ MeV. These cuts  are again chosen to match those of LSND's  $\nu e$ elastic scattering analysis \cite{lsnd_elastic}. Finally $\De = \de p_{\pi_0} \de \theta \de \ph/(2\pi)$ reflects the step size in solid angle and pion momentum, corresponding to each simulated trajectory.

\begin{figure}
\centerline{\includegraphics[width=0.75\textwidth]{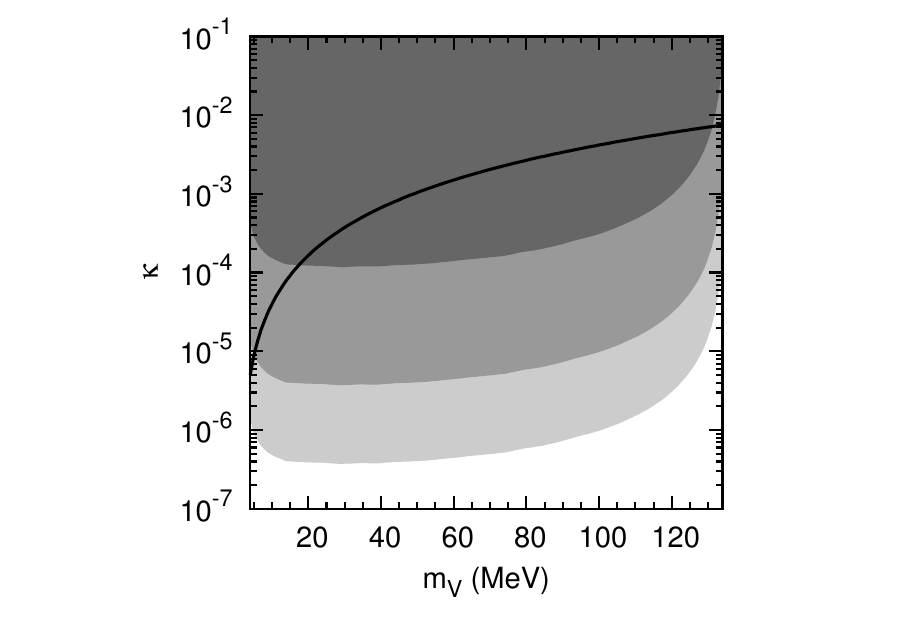}}
 \caption{\footnotesize Expected number of elastic scattering events of dark matter off electrons at the LSND detector for $m_\chi=1$ MeV. The regions show greater than 10 (light), 1000 (medium) and $10^6$ (dark) expected events. The area below the black line corresponds to $\alpha'>4 \pi$.}
\label{fig:LSND1pion} 
\end{figure}

The results of the simulation, for a range of values of $m_V$ up to the pion threshold are shown in
Fig. \ref{fig:LSND1pion}, where we have plotted the expected number of events (from (\ref{eq:Nevents})) that would be detected by LSND in the $\kappa$-$m_V$ parameter space for dark matter with $m_\chi=1$ MeV. The sensitivity is consistent with the earlier analysis in \cite{bpr99c}. In addition, the solid black line delineates the strong coupling boundary, below which $\alpha'>4\pi$. While this marks the regime where our perturbative calculations cease to be reliable, constraints on self-interaction can be somewhat stronger than this. LSND only observed ${\cal O}(300)$ beam-on events passing their cuts, of which around 200
were expected to be elastic scattering events due to neutrinos \cite{lsnd_elastic}. Thus, in light of the fact that DM scattering events should pass the same cuts, we can easily exclude the medium and dark regions of Fig. \ref{fig:LSND1pion}, for which LSND should have observed far in excess of a few hundred events. These limits could be further improved by a spectral analysis of the recoiling electrons, as the energy of the dark matter beam is considerably higher than that expected for the neutrinos from pion and muon decays at rest, and correspondingly recoil electrons from dark matter scattering would be more energetic than those from the neutrino beam. With such an analysis, one could plausibly compare the expected number of events with those due to the far rarer neutrinos produced by pion decays in flight, of which LSND observed ${\cal O}(10)$. However, even without the extra sensitivity that spectral information would provide, we can exclude the majority of the parameter space for $10~{\rm MeV}< m_V<m_\pi$ (for $m_\ch \sim 1~$MeV) through a combination of the  strong coupling condition and the expected number of NC-like scattering events. We will next look at the sensitivity of the MiniBooNE experiment, which can probe parts of the model parameter space with larger $m_\chi$ and $m_V$.

\subsection{Dark matter beams at MiniBooNE}
\label{ssec:DMMini}
The MiniBooNE experiment made use of an 8.9 GeV proton beam impacting a Be target \cite{miniflux}, and thus can produce a more energetic dark matter beam. The
process for calculating the number of neutral current-like elastic scattering events is quite similar to that described for LSND in section \ref{ssec:DMLSND}, though with some differences as outlined below. 

As for LSND, the number of $\pi^0$'s produced, $N_{\pi^0}$, was approximated by the number of charged pions produced over the lifetime of the experiment, $N_\pi^+$. As most of these pions do not decay at rest, we account for the forward boost by replacing $d\Omega_{\textnormal{lab}}$ in (\ref{eq:Npi}) with the fractional solid angle in the pion centre of mass frame, $d\Omega_{\textnormal{cm}}\simeq\gamma^2 d\Omega_{\textnormal{lab}}$ for small angles. The average $\pi^+$ energy was 1.12 GeV \cite{miniflux}, which corresponds to $\gamma \simeq 8$. Using $A_{\textnormal{det}} \simeq 1.2 \times 10^{6}$ cm$^2$, $\Phi_\nu=3.35 \times 10^{11}\,\nu\,\textnormal{cm}^{-2}$ and $(d\Omega_{\textnormal{lab}}/4\pi)_\nu=3.2 \times 10^{-5}$,  leads to $N_{\pi^+} \simeq 1.6 \times 10^{20}$. However, this is a significant over-estimate, due to the influence of a magnetic focusing horn, used to allow the experiment to run in either neutrino or anti-neutrino mode. The influence of the horn is easy to isolate, though, as the beam was run for a short period with the horn turned off. The neutrino flux was observed to drop by a factor of six during this period \cite{mininucl}, and we decrease the value of $N_{\pi^+}$ accordingly, arriving at the following estimate for $\pi^0$ production at MiniBooNE: $N_{\pi^0} \approx 2.6 \times 10^{19}$.

It is apparent that the number of neutral mesons produced at MiniBooNE is nearly three orders of magnitude lower than the number at LSND, but this is compensated to a significant extent by the large forward boost that the beam acquires, which tends to enhance the number of $\ch$'s whose trajectories intersect the detector.  Moreover, 
the MiniBooNE proton beam is of considerably higher energy than that of LSND, and as such is capable of producing $\eta$'s in significant quantities. With a mass of 547.8 MeV, including $\eta$ production allows us to greatly extend the range of $V$ masses that can be probed using these fixed target neutrino experiments. In addition, larger values of $m_\ch$ are also accessible, and we have calculated MiniBooNE's sensitivity to $m_\chi=50$ MeV in addition to $m_\chi=1$ MeV. To estimate the
$\et$ production rate, we make use of some early experimental data \cite{etacross}, which indicates that in the appropriate energy range,
\be
  \sigma_{pp \to pp\pi} \approx 30\, \sigma_{pp \to pp\eta}.
\ee
We use this ratio to normalize the number of $\et$'s produced over the lifetime of the experiment to $N_{\pi^0}$, and arrive at an estimate of $N_\eta \approx 9 \times 10^{17}$. In order to determine the sensitivity of MiniBooNE to the model, we will combine the results for dark matter from both $\eta$ and $\pi^0$ decays.

The Monte Carlo simulation follows similar lines to that described for LSND. The normalized distributions for $\pi^0$ and $\et$ production in this case were approximated by averaging the Sanford and Wang fits for $\pi^+$ and $\pi^-$ production used by MiniBooNE \cite{miniflux}, which we denote $f^{\rm SW}_{\pi^0}(\theta,p) \sim f^{\rm SW}_\et(\theta,p)$. The number of expected electron scattering events then takes the schematic form,
\be
\label{NeMB}
N^{\rm MB}_{e\chi \to e\chi} = n_e  \times \sum_{\ph=\pi^0,\et} N_{\ph} \textnormal{Br}_{\ph \to \gamma V} \times \sum_i L_i \sigma_{e\chi\to e\chi}(E_i) f_{\ph}^{\rm SW}(\theta_i,p_i) \Delta_i,
\ee
where we have used the same notation as (\ref{eq:Nevents}), and for MiniBooNE $n_e = 5.1 \times 10^{23}$ electrons/cm$^{3}$. $N_{\pi^0}$ and $N_\eta$ are given
above and provide the overall normalization,  while the electron scattering cross section is given in (\ref{eq:Nevents}). Ultimately, we find that the ability to probe the model through neutral current-like elastic scattering events with electrons is still somewhat weaker than LSND for $m_V$ below the pion threshold, but crucially it can extend the sensitivity range for $m_V$ up to $m_\eta$. We should note that there is currently no published experimental analysis for elastic scattering with electrons at MiniBooNE, so while we will determine the potential sensitivity of the experiment, this will be overly optimistic as there are no cuts imposed on the electron recoil
and we have ignored other efficiency factors.

\begin{figure}[t]
  \centerline{\includegraphics[width=0.73\textwidth]{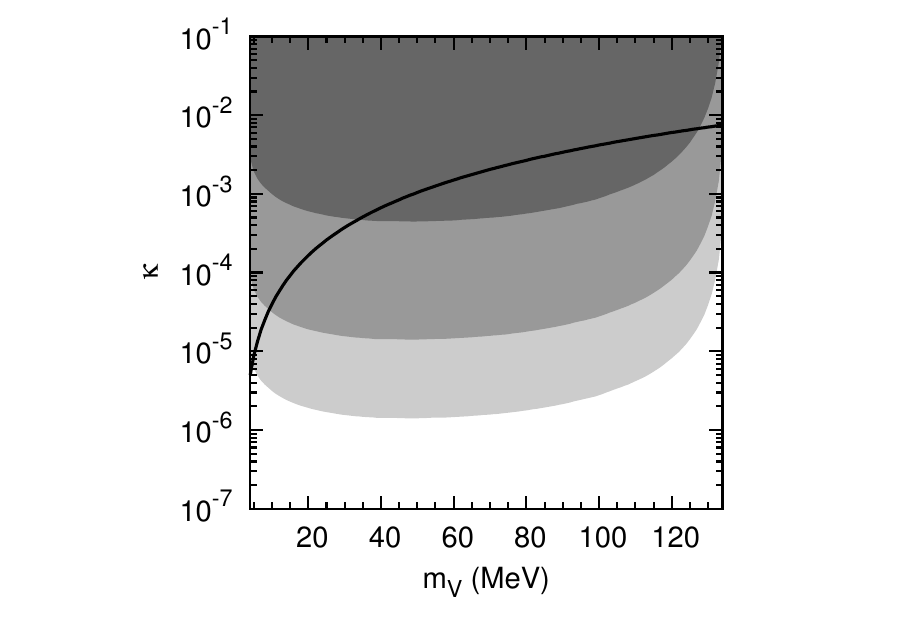} \hspace*{-3.8cm} \includegraphics[width=0.73\textwidth]{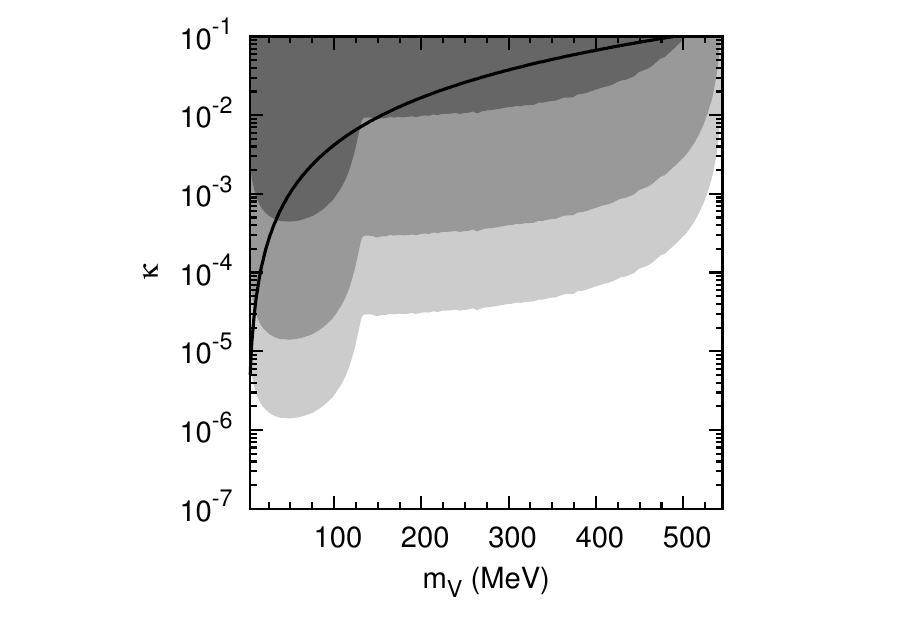}}
  \caption{\footnotesize Expected number of neutral current-like dark matter electron scattering events at the MiniBooNE detector for $m_\chi=1$ MeV. The regions show greater than 10 (light), 1000 (medium) and $10^6$ (dark) expected events. The plot on the left shows dark matter resulting from $\pi^0$ decays, while the plot on the right combines dark matter from both $\pi^0$ and $\eta$ decays. The area below the black line corresponds to $\alpha'>4 \pi$.}
  \label{fig:Mini1}
\end{figure}

In order to provide a direct comparison between the sensitivity of MiniBooNE and LSND, we first present the estimated number of neutral current-like elastic dark matter electron scattering events for $m_\chi=1$ MeV dark matter produced via pion decays in the left panel of Fig. \ref{fig:Mini1}. The shape of this plot is very similar to that found in Fig. \ref{fig:LSND1pion}, but MiniBooNE's sensitivity is down by an order of magnitude, even in the absence of cuts. However, in the right panel we present a similar plot incorporating the contribution from $\eta$ decays. The bump in $N_{\textnormal{events}}$ at low $m_V$ represents the pion contribution, and while it drops by about an order of magnitude for higher masses, the plot clearly illustrates the utility of the MiniBooNE dataset in providing sensitivity all the way up to $m_V \sim 0.5$~GeV.

While there is currently no published analysis of electron elastic scattering, MiniBooNE has recently published a full analysis of NCE scattering off nucleons\cite{mininucl}, which may lead to a substantial boost in sensitivity relative to electron scattering for certain mass regimes. In order to utilize this channel, we require the differential cross-section for neutral current-like elastic scattering between dark matter and nucleons (see Fig. \ref{f1}). This process is somewhat similar to the (vector part of) $Z$-mediated neutrino-nucleon elastic scattering (see e.g. \cite{nucleonscatter}),  and we obtain the following differential cross section,
\begin{equation}
\label{eq:nscatter}
\frac{d\sigma_{\ch N \to \ch N}}{dE_\ch} = \frac{\alpha' \kappa^2}{\alpha} \times \frac{4 \pi \alpha^2 \left[F^2_{1,N}(Q^2)A(E,E_\ch)-\frac{1}{4}F^2_{2,N}(Q^2)B(E,E_\ch)\right]}{{\left(m_V^2 + 2 m_N (E - E_\ch)\right)^2 (E^2-m_\chi^2)}},
\end{equation}
where $E$ and $E_\ch$ are the energies of the incident and outgoing dark matter particles, respectively and $Q^2=2m_N (E-E_\ch)$ is the momentum transfer.
The functions $A$ and $B$ are defined as:
\begin{align}
A(E,E_\ch) & =2 m_N E E_\ch - m_\chi^2 (E-E_\ch),  \label{eq:A} \\
B(E,E_\ch) & =(E_\ch-E)\left[(E_\ch+E)^2+2 m_N(E_\ch-E)-4m_\chi^2\right]. \label{eq:B}
\end{align}
The cross section holds for both neutrons and protons so long as the appropriate nuclear form factors are used. We make use of the following monopole and dipole form factors
\be
  F_{1,N} = \frac{q_N}{(1+Q^2/m_N^2)^2}, \;\;\;\; F_{2,N} = \frac{\kappa_N}{(1+Q^2/m_N^2)^2},
\ee
where $q_p=1$, $q_n=0$, while $\kappa_p=1.79$ and $\kappa_n=-1.9$. Equation (\ref{eq:nscatter}) only describes the scattering cross section for free nucleons, but in reality, dark matter will be scattering  off nucleons bound in a carbon nucleus or one of the two hydrogen nuclei of a CH$_2$ molecule. Following the MiniBooNE analysis \cite{mininucl}, we write the effective differential cross section as follows,
\begin{equation}
\label{eq:ch2scatter}
 \frac{d \sigma^{\rm eff}_{\chi N \to \chi N}}{dE_\ch}=\left[\frac{1}{7} C_{pf}(Q^2) + \frac{3}{7} C_{pb}(Q^2)\right]\frac{d\sigma_{\ch p \to \ch p}}{dE_\ch}+\frac{3}{7} C_{nb}(Q^2)\frac{d\sigma_{\ch n \to \ch n}}{dE_\ch},
\end{equation}
where the $C$'s describe MiniBooNE's relative efficiencies for detecting scattering off one of the protons or neutrons bound in a carbon molecule or one of the protons making up the hydrogen nuclei. The efficiencies are dependent on the momentum transfer of the scattering, but are quite close to unity for $Q^2 \in [0.4, 1]$~GeV$^2$ \cite{mininucl}. We have not made a distinction between bound and free proton scattering cross sections in (\ref{eq:ch2scatter}) or the numerical analysis.

The final expression for the expected number of NCE-like nucleon dark matter scattering events at MiniBooNE is very similar to (\ref{NeMB}). Approximating the chemical composition of the mineral oil used in the detector as pure CH$_2$, we obtain
\be
\label{eq:NeventsNucleon}
N^{\rm MB}_{N\chi \to N\chi} = 14 n_{\textnormal{CH}_2} \times \ep_{\rm eff} \times \sum_{\ph=\pi^0,\et} \left( N_{\ph} \textnormal{Br}_{\ph \to \gamma V}  \sum_i L_i \sigma^{\rm eff}_{N\chi\to N\chi}(E_i) f_{\ph}^{\rm SW}(\theta_i,p_i) \Delta_i \right),
\ee
where $n_{\textnormal{CH}_2}= 6.4\times 10^{22}$ molecules/cm$^{3}$ is the number density of CH$_2$ in the mineral oil used at MiniBooNE, while $\ep_{\rm eff} \simeq 0.59$ is the detection efficiency \cite{perevalov} for events within the specific fiducial volume and momentum transfer cuts imposed in the MiniBooNE NCE analysis  \cite{mininucl}, that we adopt here to allow for a direct comparison. The momentum transfer cut of 0.1 -- 1.6 GeV$^2$ determines the range over which (\ref{eq:ch2scatter}), which weights over proton and neutron scattering, is integrated to produce the effective cross-section 
$\sigma^{\rm eff}_{N\chi\to N\chi}$. Note that the lower cut at 0.1 GeV$^2$ means that there is no sensitivity to coherent nuclear elastic scattering, and our nucleon-level treatment should be reliable. The remaining factors in (\ref{eq:NeventsNucleon}) are defined as in (\ref{NeMB}).

\begin{figure}[t]
\centerline{\includegraphics[width=0.73\textwidth]{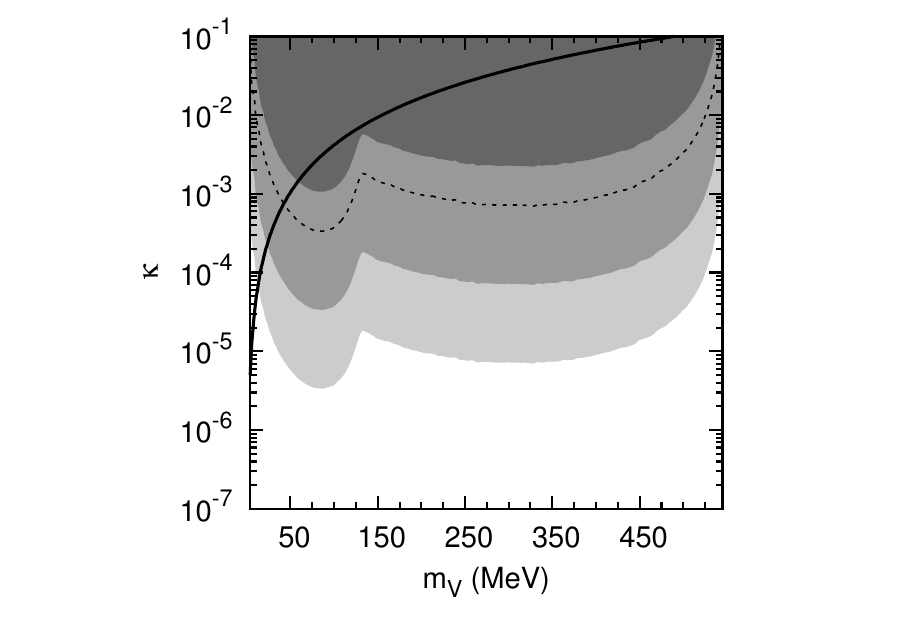} \hspace*{-3.8cm} \includegraphics[width=0.73\textwidth]{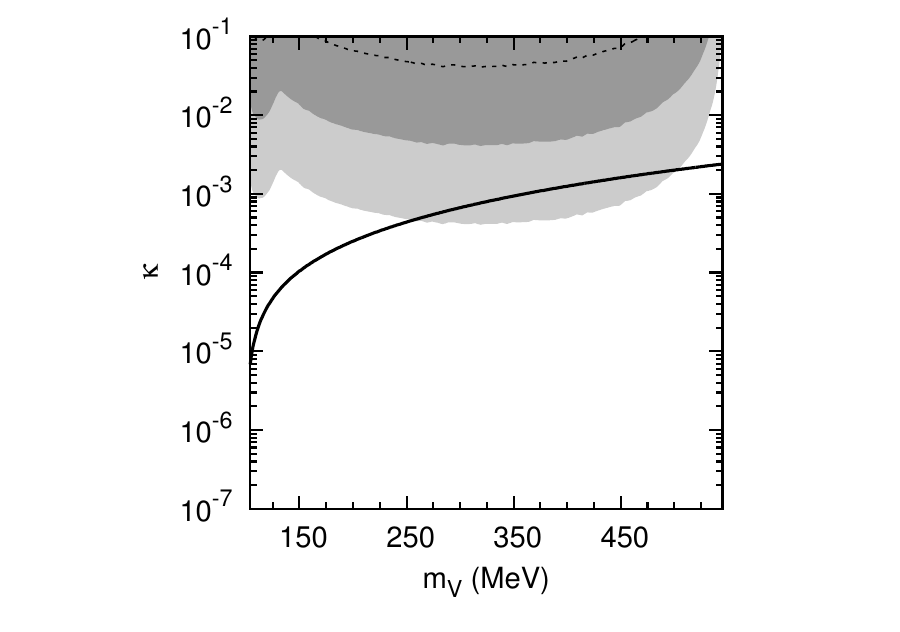}}
  \caption{\footnotesize Expected number of neutral current-like dark matter nucleon scattering events at the MiniBooNE detector.  The regions show greater than 10 (light), 1000 (medium) and $10^6$ (dark) expected events. The plot on the left is for $m_\chi=1$ MeV, while the plot on the right is for $m_\chi=50$ MeV. The area below the black line corresponds to $\alpha'>4 \pi$, while the dashed curve indicates the total number of (background) neutrino events observed.}
  \label{fig:Mini2}
\end{figure} 
 
The results for DM nucleon scattering at MiniBooNE are shown in the left panel of Fig. \ref{fig:Mini2}. In contrast to the situation for NCE neutrino scattering, the nucleon scattering sensitivity here is slightly weaker than for electron scattering at low $V$ masses, though it does improve as we increase the $V$ mass.  This apparent suppression of the nucleon scattering cross-section  can be understood as follows. The usual NCE-type enhancement naturally emerges here in the $m_V^2 \gg Q^2$ limit, where the relative nucleon vs electron scattering cross-sections scale as $\si_{N\ch\rightarrow N\ch}/\si_{e\ch\rightarrow e\ch} \sim m_N/m_e \sim 10^3$. In contrast, the characteristic regime covered here is $Q^2 > m_V^2$ and this large enhancement for nucleon scattering is lost. Nonetheless, we see that as $m_V$ increases the 
sensitivity increases, and actually peaks around $m_V \sim$ 300 MeV for events due to $\et$ decays. 
In practice, fully exploiting this strong underlying sensitivity at MiniBooNE would require a careful analysis of the spectral information, in order to isolate DM scattering events within the total elastic scattering dataset which contains approximately $9.5\times 10^4$ events (of which roughly 65\% are expected to be actual NCE scatterings) \cite{mininucl}. While all these events are necessarily similar, and we rely on this to conclude that DM scattering would pass the signal cuts, the characteristic DM beam energy should differ from the characteristic energy of the neutrinos, so separating the signal
from the neutrino `background' seems feasible at some level of precision. However, even without a more sophisticated analysis of this type, the fact that the strong coupling condition $\al' < 4\pi$ becomes increasingly restrictive at higher $V$ masses, allows us to rule out almost the entire parameter space of 1 MeV dark matter for $2m_\ch < m_V<m_\eta$, particularly when these results are combined with the exclusion region from LSND.

Independent of any motivation for such models as explanations of the galactic 511~keV line, it is interesting to explore how far the sensitivity reach extends in DM mass. It is in this spirit that we show the result of repeating the previous analysis for dark matter with $m_\chi=50$ MeV in the right-hand plot of Fig. \ref{fig:Mini2}. We find that there is still significant sensitivity of MiniBooNE to this dark matter mass range, while the bounds imposed by the strong coupling condition are somewhat weaker. Thus a significant portion of the parameter space is still allowed by experimental data, under the conservative assumption that the ${\cal O}(10-1000)$ events for the parameter region above the strong coupling boundary could be hiding in the large NCE dataset. While not shown here, this calculation was repeated for $m_\chi=100$ MeV leading to similar, though slightly weaker, bounds.

\section{Discussion and Conclusions}
\label{sec:Conc}

There has been a recent resurgence of interest in MeV-to-GeV scale phenomenology, in which considerable attention has been 
devoted to the search for light mediator particles. If such mediators are the lightest members of an extended 
hidden (or dark) sector, it is then guaranteed that if produced they will at some point decay back to the SM. An explicit example
is a new light U(1)$'$ vector coupled to the SM via the vector portal. A number of search strategies are based on the production 
of such a vector with its subsequent decay to electron-positron pairs (see {\em e.g.} \cite{best}). However, these search strategies are severely limited
if, instead, the mediator is not the lightest hidden sector state and decays predominantly to some other light states. In this case, the scattering of those light states in a detector spatially separated from the production point represents perhaps the most efficient search strategy. Moreover, owing to the potentially large production rate, and the existence of large volume detectors, proton fixed-target facilities focusing on neutrino physics appear to be an ideal means for exploring these scenarios.
In this paper, we have given a detailed example of how such a search could be undertaken, with 
MeV-scale DM models as our primary focus. In turn, we have found that the most natural/economical model realization 
of TRDM in the MeV range -- with vector portal mediation -- is under severe pressure from the experimental constraints imposed by the absence of any non-standard
elastic scattering signal at MiniBooNE and LSND. In particular, in the limited mass range $m_\ch \sim 1-3$~MeV that is required by other $\gamma$-ray limits \cite{by}
for any putative explanation of the galactic 511~keV line, these constraints rule out this candidate for most values of the vector mass below the 
$\eta$ threshold. 

In the remainder of this section, we will comment briefly on other sources of constraints
on these models, which are generally less sensitive than the fixed target probes discussed here, and also mention some future directions.

\bigskip
\noindent$\bullet$ {\it Additional particle physics constraints:}
To illustrate the advantage of constraints derived from the DM beam analysis, we can compare them with those
that follow from meson decays and precision QED measurements. The anomalous magnetic moment of the muon is 
a very sensitive probe of sub-GeV physics (see {\em e.g.} \cite{Fayet, tests}). In the vector portal scenario studied here, 
additional positive contributions to $g-2$ originate from a muon-mediator loop, and the current sensitivity corresponds to 
mixing angles  in the range $\kappa \sim 10^{-3}-10^{-2}$ \cite{tests}. From the results of this paper (Figures 2 and 4),  
one can see that {\em if} the vector mediator is unstable with respect to decays to lighter MeV-scale 
states, the constraints derived from LSND and MiniBooNE are far more stringent, and in particular 
do not allow us to ascribe the current discrepancy between the measurement and the SM calculations of $(g-2)_\mu$
to a light vector mediator with $\ka \sim {\cal O}(10^{-3})$.

The vector isosinglet nature of the minimal vector portal provides a natural defense against excessive
flavor-changing effects in the meson sector. Nonetheless, the two-body decay $K^+\to \pi^+V$ followed by the invisible 
decay of the mediator will give an additional contribution to the tiny, but measured, branching fraction $K^+ \rightarrow \pi^+ E\!\!\!/$. 
Using the results of \cite{tests}, we estimate that the maximal contribution 
of the vector mediator to the decay rate cannot exceed ${\rm Br}_{K\to \pi V \to \pi E\!\!\!/} \sim 
3 \times 10^{-10} \times ( \kappa/10^{-3})^2$. Given that this decay  is observed with about this branching 
ratio, we conclude that the sensitivity to $\kappa$ does not exceed $10^{-3}$, which is again inferior to the constraints
derived in this paper. Finally, direct experimental constraints on the missing energy events 
$e^+e^-\to \gamma V\to \gamma E\!\!\!/$ at B-factories were discussed in \cite{Drees}. 
The sensitivity of this process to $\kappa$ currently does not exceed $10^{-2}$, as such a search 
is very difficult to implement.

\bigskip
\noindent$\bullet$ {\it Astrophysical and cosmological constraints:}
We should also address the question of astrophysical constraints on light mediators and MeV-scale DM. As we have discussed, the vector 
portal model considered here naturally produces a galactic 511~keV flux below or approaching the observations by
INTEGRAL/SPI, but there are other possible sources of astrophysical or cosmological constraints. 
For example, in the early universe the lower end of the mass range for MeV DM implies that it will freeze-out quite close
to the BBN epoch. The annihilation products can then have an effect on the light element abundances. This question
was analyzed in \cite{raffelt}, with the result that this effect is rather small for $m_\ch \geq 1$ MeV. Another 
possible source of constraints is the impact of DM annihilation on the CMB in the late universe \cite{CMBa}. Such constraints
can be quite stringent for MeV-scale DM that has an $s$-wave annihilation rate, as the annihilation products have energies
in an ideal range to heat the IGM. Nonetheless, in the present case these constraints disappear as the $p$-wave annihilation
is highly suppressed in this epoch where DM is very cold.

A potentially promising direct source of (semi-)relativistic MeV-scale WIMPs are certain extreme astrophysical 
environments. Indeed, the maximum attainable energies in the core of supernova explosions are in excess of 10-20 MeV. 
However, given that the interaction rate of MeV dark matter with the electrons is appreciably higher than the weak 
rate, as is the case for the parameter range studied here, any DM states produced will be thermalized and trapped in the core, and 
thus will not lead to additional constraints via new energy loss mechanisms. Given its coupling to the SM via the vector and/or Higgs portals, it
is also guaranteed that the neutrino spectrum is not degraded and remains consistent with the energy range suggested by the 
detection of the SN1987a neutrino signal. To escape the core, DM would have to diffuse the same way neutrinos do. However, while 
the neutrino energy spectra can be characterized as almost thermal with a `temperature' in the range of 5 to 10 MeV, the corresponding 
 MeV DM `temperature' will have to be smaller due to its larger interaction rate with electrons and positrons. It is then guaranteed that 
 the SN freeze-out `temperature' for $\chi$ will be $T_\chi < m_e$. A population of energetic light DM particles created by past SN explosions is 
 in principle detectable via their interaction with {\em e.g.} electrons and residual ionization. However, given that the diffuse SN 
 neutrino background still remains undetected, discovering MeV scale particles this way represents a serious technical challenge.

\bigskip
\noindent$\bullet$ {\it Future progress:}
One interesting avenue for extending the sensitivity reach in the low mass range would be to 
exploit any future experimental facilities aiming to measure the coherent neutrino-nucleus elastic cross-section. 
Indeed, it is well known that a source of stopped pions can be used to detect the
elastic  scattering of neutrinos on nuclei in  DM-type detectors, sensitive to a recoil energy in the ${\cal O}(10~{\rm keV})$ 
range. Should MeV DM exist, it would also produce a considerable recoil, possibly
dominating over the coherent neutrino scattering signal. Such a possibility provides some additional physics motivation for the 
proposals such as CLEAR \cite{clear}, that are capable of detecting the coherent scattering of neutrinos. 
Finally, we note that modern long-baseline neutrino facilities such as MINOS and T2K have more energetic beams and may open the
possibility of extending the mass reach at the upper end.

\section*{Note Added [August 12, 2013]}

In the two years since the publication of this paper, there has been growing interest in the use of fixed target facilities to probe scenarios of light sub-GeV dark matter. In this short note, we attempt to clarify the reach of LSND's analysis of neutrino-electron elastic scattering \cite{lsnd_elastic} with regard to the parameter space of the vector-mediated dark matter model. Event contours already appear in Fig.~2, and our aim is simply to infer the contour representing the 90\% confidence level limit.  As noted in Sect.~3.1, LSND observed only 301 beam-on events in the dataset, of which 242 were identified as elastic scattering by neutrinos, independent of flavor. This compares to the Standard Model background of 229. Given the uncertainties, at 90\% confidence there were less than 55 non-standard scattering events \cite{lsnd_elastic}, even before detailed consideration of the kinematics. The most significant theoretical uncertainty is the overall normalization of the $\pi^0$ production rate. We will assign a factor of 2 uncertainty to this rate which, combined with the 90\% confidence limit of 55 scattering events, leads to the contour shown in Fig.~5. In practice, the sensitivity to the overall production rate scales as $\kappa \propto (N_{\pi^0})^{1/4}$ and thus is quite mild. The analysis of the event rate is as discussed in Sect.~3, but in contrast to Fig.~2 this plot does not impose the restriction that $\al'$ is fixed to ensure the correct relic abundance. Instead, we fix $\al'=\al$ (the number of events scales linearly with $\al'$), and show the relic density curve separately, allowing this limit to more easily be scaled for other models in which, for example, this candidate only makes up a fraction of the full dark matter abundance. The plot also shows the indirect limits from corrections to muon and electron $g-2$ values, and the band that is favored to bring $g-2$ for the muon into line with experiment \cite{tests}.

\begin{figure}[t]
\centerline{\includegraphics[width=0.48\textwidth]{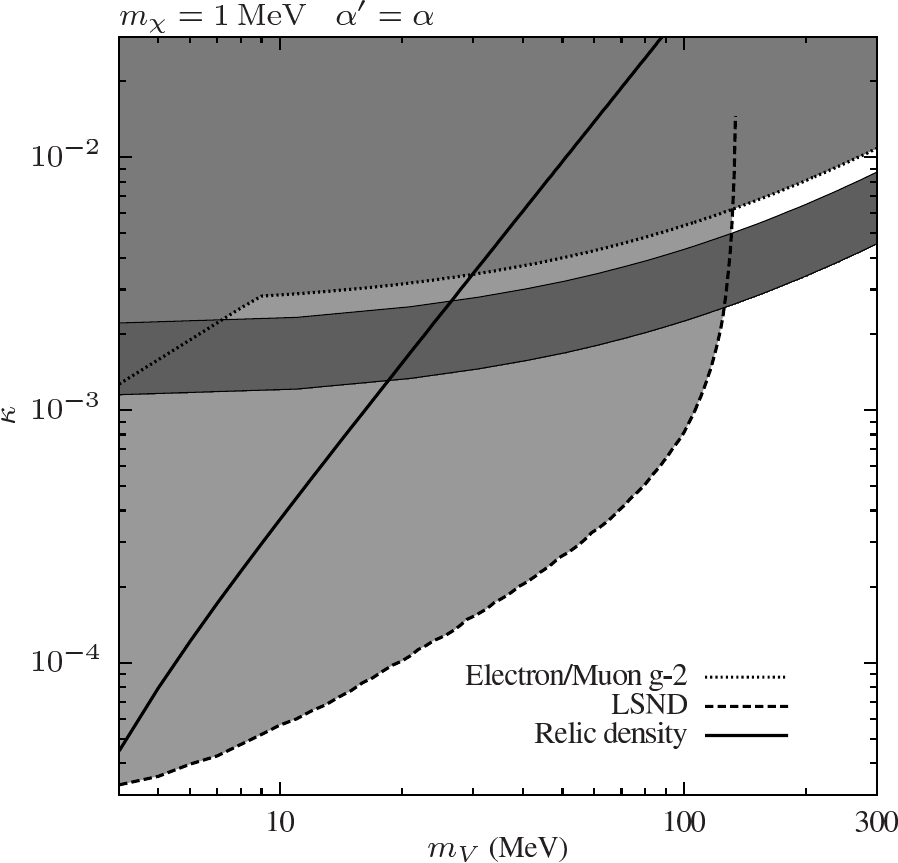} \hspace*{0.4cm} \includegraphics[width=0.48\textwidth]{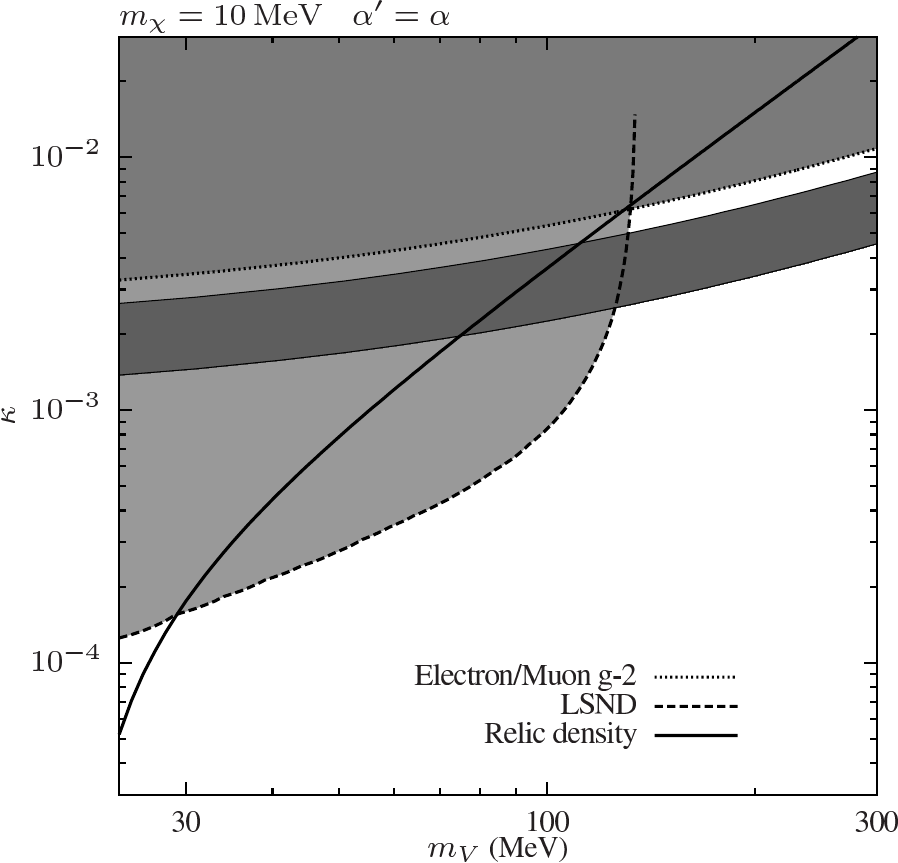}}
  \caption{\footnotesize As in Fig.~2, we show the sensitivity of LSND's elastic scattering analysis \cite{lsnd_elastic} to scattering of dark matter off electrons, when $m_\chi< m_V/2$. The dashed LSND contour corresponds to the 90\% confidence limit as discussed in the text, based on LSND's comparison of the observed data to the Standard Model neutrino background \cite{lsnd_elastic}. The confidence limit is shown for two dark matter masses, 1~MeV (left) and 10~MeV (right). The solid black line indicates the parameters required to ensure saturation of the dark matter relic density. The darker shaded regions show exclusions due to loop corrections from the vector to $g-2$ of the muon and electron, while the dark band is the preferred region to shift the muon $g-2$ value into line with experiment \cite{tests}. }
\end{figure} 

\section*{Acknowledgements}

We would like to thank Brian Batell for numerous helpful discussions and correspondence.
The work of P.dN. was supported in part by a Canada Graduate Scholarship from NSERC, Canada. The work of  M.P. and A.R. was also supported in part by NSERC, Canada, and research at the Perimeter Institute
is supported in part by the Government of Canada through NSERC and by the Province of Ontario through MEDT.


\begin{thebibliography}{99}


\bibitem{Boehm}
  C.~Boehm, D.~Hooper, J.~Silk, M.~Casse and J.~Paul,
  %``MeV dark matter: Has it been detected?,''
  Phys.\ Rev.\ Lett.\  {\bf 92}, 101301 (2004)
  [arXiv:astro-ph/0309686].

 \bibitem{Fayet}
  P.~Fayet,
  %``Light spin-1/2 or spin-0 dark matter particles,''
  Phys.\ Rev.\ D {\bf 70}, 023514 (2004)
  [arXiv:hep-ph/0403226];  P.~Fayet,
  %``Constraints on light dark matter and U bosons, from psi, Upsilon, K+, pi0,
  %eta and eta' decays,''
  Phys.\ Rev.\  D {\bf 74}, 054034 (2006)
  [arXiv:hep-ph/0607318];
  P.~Fayet,
  %``U-boson production in e+ e- annihilations, psi and Upsilon decays, and
  %light dark matter,''
  Phys.\ Rev.\  D {\bf 75}, 115017 (2007)
  [arXiv:hep-ph/0702176].

  
   \bi{BF} C.~Boehm and P.~Fayet,
  %``Scalar dark matter candidates,''
  Nucl.\ Phys.\  B {\bf 683}, 219 (2004)
  [arXiv:hep-ph/0305261].



\bi{light-chi} 
  P.~Gondolo and G.~Gelmini,
  %``Compatibility of DAMA dark matter detection with other searches,''
  Phys.\ Rev.\ D {\bf 71}, 123520 (2005)
  [arXiv:hep-ph/0504010].
  %%CITATION = HEP-PH 0504010;%%
  

\bi{Neil}
  D.~P.~Finkbeiner and N.~Weiner,
  %``Exciting Dark Matter and the INTEGRAL/SPI 511 keV signal,''
  Phys.\ Rev.\  D {\bf 76}, 083519 (2007)
  [arXiv:astro-ph/0702587].

  
\bi{PRV}  M.~Pospelov, A.~Ritz and M.~B.~Voloshin,
  %``Secluded WIMP Dark Matter,''
  Phys.\ Lett.\  B {\bf 662}, 53 (2008)
  [arXiv:0711.4866 [hep-ph]].

\bi{HZ} D.~Hooper and K.~M.~Zurek,
  %``Natural supersymmetric model with MeV dark matter,''
  Phys.\ Rev.\  D {\bf 77}, 087302 (2008)
  [arXiv:0801.3686 [hep-ph]].


 
\bi{AFSW}
  N.~Arkani-Hamed, D.~P.~Finkbeiner, T.~Slatyer and N.~Weiner,
  %``A Theory of Dark Matter,''
  Phys.\ Rev.\  D {\bf 79}, 015014 (2009)
  [arXiv:0810.0713 [hep-ph]].
  %%CITATION = ARXIV:0810.0713;%%

\bi{PR}
  M.~Pospelov and A.~Ritz,
  %``Astrophysical Signatures of Secluded Dark Matter,''
  Phys.\ Lett.\  B {\bf 671}, 391 (2009)
  [arXiv:0810.1502 [hep-ph]].


  \bi{Drees} N.~Borodatchenkova, D.~Choudhury and M.~Drees,
  %``Probing MeV dark matter at low-energy e+ e- colliders,''
  Phys.\ Rev.\ Lett.\  {\bf 96}, 141802 (2006)
  [arXiv:hep-ph/0510147].




\bi{tests}
  M.~Pospelov,
  %``Secluded U(1) below the weak scale,''
  Phys.\ Rev.\  D {\bf 80}, 095002 (2009)
  [arXiv:0811.1030 [hep-ph]].
  
\bi{BPR}  B.~Batell, M.~Pospelov and A.~Ritz,
  %``Probing a Secluded U(1) at B-factories,''
  Phys.\ Rev.\ D {\bf 79}, 115008 (2009)
  [arXiv:0903.0363 [hep-ph]].

\bi{slac}  R.~Essig, P.~Schuster and N.~Toro,
  %``Probing Dark Forces and Light Hidden Sectors at Low-Energy e+e-
  %Colliders,''
  Phys.\ Rev.\  D {\bf 80}, 015003 (2009)
  [arXiv:0903.3941 [hep-ph]].


\bi{Reece} M.~Reece and L.~T.~Wang,
  JHEP {\bf 0907}, 051 (2009)
  [arXiv:0904.1743 [hep-ph]].
  
  \bi{best}
  J.~D.~Bjorken, R.~Essig, P.~Schuster, N.~Toro,
  %``New Fixed-Target Experiments to Search for Dark Gauge Forces,''
  Phys.\ Rev.\  {\bf D80}, 075018 (2009).
  [arXiv:0906.0580 [hep-ph]].

  
  \bi{bpr99c}
B.~Batell, M.~Pospelov, A.~Ritz,
  %``Exploring Portals to a Hidden Sector Through Fixed Targets,''
  Phys.\ Rev.\ D {\bf 80}, 095024 (2009).
  [arXiv:0906.5614 [hep-ph]].

\bi{others}
P.~Schuster, N.~Toro, I.~Yavin,
  %``Terrestrial and Solar Limits on Long-Lived Particles in a Dark Sector,''
  Phys.\ Rev.\  {\bf D81}, 016002 (2010).
  [arXiv:0910.1602 [hep-ph]]; 
  R.~Essig, P.~Schuster, N.~Toro, B.~Wojtsekhowski,
  %``An Electron Fixed Target Experiment to Search for a New Vector Boson A' Decaying to e+e-,''
  JHEP {\bf 1102}, 009 (2011).
  [arXiv:1001.2557 [hep-ph]];
R.~Essig, R.~Harnik, J.~Kaplan, N.~Toro,
  %``Discovering New Light States at Neutrino Experiments,''
  Phys.\ Rev.\  {\bf D82}, 113008 (2010).
  [arXiv:1008.0636 [hep-ph]].


\bi{losecco}  J.~LoSecco {\it et al.},
  %``Limits On The Production Of Neutral Penetrating States In A Beam Dump,''
  Phys.\ Lett.\  B {\bf 102}, 209 (1981).



  \bi{pw}
B.~Patt and F.~Wilczek,
  %``Higgs-field portal into hidden sectors,''
  arXiv:hep-ph/0605188.
  %%CITATION = HEP-PH/0605188;%%
  
  \bi{portal1}
  R.~Foot, H.~Lew and R.~R.~Volkas,
  %``A Model with fundamental improper space-time symmetries,''
  Phys.\ Lett.\  B {\bf 272}, 67 (1991).
  %%CITATION = PHLTA,B272,67;%%
  R.~Foot and X.~G.~He,
  %``Comment On Z Z-Prime Mixing In Extended Gauge Theories,''
  Phys.\ Lett.\  B {\bf 267}, 509 (1991).
  %%CITATION = PHLTA,B267,509;%%
 
 \bi{higgsportal} 
  R.~Schabinger and J.~D.~Wells,
  %``A minimal spontaneously broken hidden sector and its impact on Higgs  boson
  %physics at the Large Hadron Collider,''
  Phys.\ Rev.\  D {\bf 72}, 093007 (2005)
  [arXiv:hep-ph/0509209].
  %%CITATION = PHRVA,D72,093007;%%
  D.~G.~Cerdeno, A.~Dedes and T.~E.~J.~Underwood,
  %``The minimal phantom sector of the standard model: Higgs phenomenology  and
  %Dirac leptogenesis,''
  JHEP {\bf 0609}, 067 (2006)
  [arXiv:hep-ph/0607157];
  %%CITATION = JHEPA,0609,067;%%
  J.~R.~Espinosa and M.~Quiros,
  %``Novel effects in electroweak breaking from a hidden sector,''
  Phys.\ Rev.\  D {\bf 76}, 076004 (2007)
  [arXiv:hep-ph/0701145];
  %%CITATION = PHRVA,D76,076004;%%
  J.~March-Russell, S.~M.~West, D.~Cumberbatch and D.~Hooper,
  %``Heavy Dark Matter Through the Higgs Portal,''
  JHEP {\bf 0807}, 058 (2008)
  [arXiv:0801.3440 [hep-ph]];
  %%CITATION = JHEPA,0807,058;%%
  
  \bi{DMportal}
  M.~Ahlers, J.~Jaeckel, J.~Redondo and A.~Ringwald,
  %``Probing Hidden Sector Photons through the Higgs Window,''
  Phys.\ Rev.\  D {\bf 78}, 075005 (2008)
  [arXiv:0807.4143 [hep-ph]];
  %%CITATION = PHRVA,D78,075005;%%
  J.~L.~Feng, H.~Tu and H.~B.~Yu,
  %``Thermal Relics in Hidden Sectors,''
  JCAP {\bf 0810}, 043 (2008)
  [arXiv:0808.2318 [hep-ph]];
  %%CITATION = JCAPA,0810,043;%%
K.~Kohri, J.~McDonald and N.~Sahu,
  %``Cosmic Ray Anomalies and Dark Matter Annihilation to Muons via a Higgs
  %Portal Hidden Sector,''
  arXiv:0905.1312 [hep-ph];
  %%CITATION = ARXIV:0905.1312;%%
  J.~L.~Feng, M.~Kaplinghat, H.~Tu and H.~B.~Yu,
  %``Hidden Charged Dark Matter,''
  JCAP {\bf 0907}, 004 (2009)
  [arXiv:0905.3039 [hep-ph]].
  %%CITATION = JCAPA,0907,004;%%

\bi{axions_exp} 
see {\it e.g.} L.~J.~Rosenberg and K.~A.~van Bibber,
  %``Searches for invisible axions,''
  Phys.\ Rept.\  {\bf 325}, 1 (2000); 
  %%CITATION = PRPLC,325,1;%%
G.~Carosi and K.~van Bibber,
  %``Cavity microwave searches for cosmological axions,''
  Lect.\ Notes Phys.\  {\bf 741}, 135 (2008)
  [arXiv:hep-ex/0701025].
  %%CITATION = LNPHA,741,135;%%


\bi{neutrino_exp}  E.~Gallas {\it et al.}  [FMMF Collaboration],
  %``Search for neutral weakly interacting massive particles in the Fermilab
  %Tevatron wide band neutrino beam,''
  Phys.\ Rev.\  D {\bf 52}, 6 (1995); G.~Bernardi {\it et al.},
  %``FURTHER LIMITS ON HEAVY NEUTRINO COUPLINGS,''
  Phys.\ Lett.\  B {\bf 203}, 332 (1988).


\bi{gluino_exp}  J.~Adams {\it et al.}  [KTeV Collaboration],
  %``Search for Light Gluinos via the Spontaneous Appearance of pi+pi- Pairs
  %with an 800 GeV/c Proton Beam at Fermilab,''
  Phys.\ Rev.\ Lett.\  {\bf 79}, 4083 (1997)
  [arXiv:hep-ex/9709028].


\bi{unstable}  J.~Badier {\it et al.}  [NA3 Collaboration],
  %``Mass And Lifetime Limits On New Longlived Particles In 300-Gev/C Pi-
  %Interactions,''
  Z.\ Phys.\  C {\bf 31}, 21 (1986).


 

%\bi{Pamela}
 % O.~Adriani {\it et al.},
  %``Observation of an anomalous positron abundance in the cosmic radiation,''
 % arXiv:0810.4995 [astro-ph].
  

\bi{integral}
N.~Prantzos, C.~Boehm, A.~M.~Bykov, R.~Diehl, K.~Ferriere, N.~Guessoum, P.~Jean, J.~Knoedlseder {\it et al.},
  %``The 511 keV emission from positron annihilation in the Galaxy,''
  [arXiv:1009.4620 [astro-ph.HE]].

\bi{bouchet}
L.~Bouchet, J.~P.~Roques and E.~Jourdain,
  %``On the morphology of the electron-positron annihilation emission as seen by
  %SPI/INTEGRAL,''
  Astrophys.\ J.\  {\bf 720}, 1772 (2010)
  [arXiv:1007.4753 [astro-ph.HE]].
  %%CITATION = ASJOA,720,1772;%%

\bi{CMBa}  N.~Padmanabhan and D.~P.~Finkbeiner,
  %``Detecting dark matter annihilation with CMB polarization: Signatures and
  %experimental prospects,''
  Phys.\ Rev.\  D {\bf 72}, 023508 (2005)
  [arXiv:astro-ph/0503486];  T.~R.~Slatyer, N.~Padmanabhan and D.~P.~Finkbeiner,
  %``CMB Constraints on WIMP Annihilation: Energy Absorption During the
  %Recombination Epoch,''
  Phys.\ Rev.\  D {\bf 80}, 043526 (2009)
  [arXiv:0906.1197 [astro-ph.CO]].
  
  \bi{lsnd_elastic}
L.~B.~Auerbach {\it et al.} [ LSND Collaboration ],
  %``Measurement of electron - neutrino - electron elastic scattering,''
  Phys.\ Rev.\  {\bf D63}, 112001 (2001).
  [hep-ex/0101039].
  
  \bi{lsndresult}
A.~Aguilar {\it et al.}, [LSND Collaboration],
%Evidence for Neutrino Oscillations from the Observation of Electron Anti-neutrinos in a Muon Anti-Neutrino Beam
Phys.\ Rev.\ D{\bf 64}, 112007 (2001).
[arXiv:hep-ex/0104049]

\bi{etacross}
S.~Teis, W.~Cassing, M.~Effenberger, A.~Hombach, U.~Mosel, Gy.~Wolf,
%Pion-Production in Heavy-Ion Collisions at SIS energies
Z.\ Phys.\ A {\bf 356}, 421 (1997).
[arXiv:nucl-th/9609009]
%add DOI?

\bi{lsndpionpara}%How do I format a tech report? gah.
R.L.~Burman and E.S.~Smith, 
%Parameterization of Pion Production and Reaction Cross Sections at LAMPF Energies.
LA-11502-MS, DE-98-011120, UC-414, Los Alamos, 1989.

\bi{lsnddetector}
C.~Athanassopoulos {\it et al.},
%The Liquid Scintillator Neutrino Detector and LAMPF Neutrino Source,
 Nucl.\ Instrum.\ Meth A {\bf 388}, 149 (1997).
[arXiv:nucl-ex/9605002]

\bi{miniflux}
A.A.~Aguilar-Arevalo {\it et al.} [MiniBooNE Collaboration],
% The Neutrino Flux prediction at MiniBooNE
Phys.\ Rev.\ D {\bf 79}, 072002 (2009).
[arXiv:0806.1449 [hep-ex]]


\bi{mininucl}
A.A.~Aguilar-Arevalo {\it et al.} [MiniBooNE Collaboration],
%Measurement of the Neutrino Neutral-Current Elastic Differential Cross Section
Phys.\ Rev.\ D {\bf 82}, 092005 (2010).
[arXiv:1007.4730 [hep-ex]]

\bi{nucleonscatter}
L.A.~Ahrens {\it et al.},
%Measurements of neutrino-proton and antineutrino-proton elastic scattering
Phys.\ Rev.\ D {\bf 35} 785 (1987).

\bi{perevalov}
D. Perevalov, 
%`Neutrino-Nucleus Neutral Current Elastis Interaction Measurement in MiniBooNE'
PhD Thesis, University of Alabama, 2009.


\bi{by}
J.~F.~Beacom, H.~Yuksel,
  %``Stringent constraint on galactic positron production,''
  Phys.\ Rev.\ Lett.\  {\bf 97}, 071102 (2006).
  [astro-ph/0512411].


\bi{raffelt}
P.~D.~Serpico, G.~G.~Raffelt,
  %``MeV-mass dark matter and primordial nucleosynthesis,''
  Phys.\ Rev.\  {\bf D70}, 043526 (2004).
  [astro-ph/0403417].
  
  
  \bi{clear}
  K.~Scholberg, T.~Wongjirad, E.~Hungerford, A.~Empl, D.~Markoff, P.~Mueller, Y.~Efremenko, D.~McKinsey {\it et al.},
  %``The CLEAR Experiment,''
  [arXiv:0910.1989 [hep-ex]];
A.~J.~Anderson, J.~M.~Conrad, E.~Figueroa-Feliciano, K.~Scholberg, J.~Spitz,
  %``Coherent Neutrino Scattering in Dark Matter Detectors,''
  [arXiv:1103.4894 [hep-ph]].



  \end{thebibliography}
\end{document}